\RequirePackage{fix-cm}

\documentclass[twocolumn]{svjour3}
\usepackage{cite}
\usepackage{amsmath,amssymb,amsfonts}
\usepackage{algorithmic}
\usepackage{graphicx}
\usepackage{textcomp}
\usepackage{xcolor}

\usepackage{amsmath,amssymb}
\usepackage{chemfig}
\usepackage[tight,footnotesize]{subfigure}
\usepackage{graphicx}
\usepackage{textcomp}
\usepackage{color}
\usepackage{dcolumn}
\usepackage{bm}
\usepackage[numbers,comma,sort&compress]{natbib}
\usepackage{gensymb}
\usepackage{csquotes}
\usepackage{subfig}
\usepackage{caption}
\usepackage{authblk}
\usepackage{amsmath}
\usepackage{amssymb}
\usepackage{multirow}
\usepackage{hhline}
\usepackage{booktabs}
\usepackage[strings]{underscore}
\usepackage{url}
\usepackage{breakurl}
\usepackage{tabularx}
\usepackage{soul}
\usepackage{amsmath}
\usepackage{float}
\DeclareMathOperator*{\argmax}{argmax} 

\graphicspath{{images/}}
\DeclareGraphicsExtensions{.pdf,.jpeg,.png}
\patchcmd{\footnotemark}{\stepcounter{footnote}}{\refstepcounter{footnote}}{}{}

\newcommand{\R}{\mathbb{R}}

\begin{document} \sloppy

\title{Conditional $\beta$-VAE for \textit{De Novo} Molecular Generation}


\author{Ryan J. Richards$\textsuperscript{12*}$ \and Austen M. Groener$\textsuperscript{1}$}

\institute{1 De Novo AI Research \\ 2 University of Pennsylvania, School of Engineering and Applied Science \\ * Corresponding author: ryry@seas.upenn.edu}
\date{Received: date / Accepted: date}


\maketitle

\begin{abstract} ---  Deep learning has significantly advanced and accelerated \textit{de novo} molecular generation. Generative networks, namely Variational Autoencoders (VAEs) can not only randomly generate new molecules, but also alter molecular structures to optimize specific chemical properties which are pivotal for drug-discovery. While VAEs have been proposed and researched in the past for pharmaceutical applications, they possess deficiencies which limit their ability to both optimize properties and decode syntactically valid molecules. We present a recurrent, conditional $\beta$-VAE which disentangles the latent space to enhance \textit{post hoc} molecule optimization. We create a mutual information driven training protocol and data augmentations to both increase molecular validity and promote longer sequence generation. We demonstrate the efficacy of our framework on the ZINC-250k dataset, achieving SOTA unconstrained optimization results on the penalized LogP (pLogP) and QED scores, while also matching current SOTA results for validity, novelty and uniqueness scores for random generation. We match the current SOTA on QED for top-3 molecules at 0.948, while setting a new SOTA for pLogP optimization at 104.29, 90.12, 69.68 and demonstrating improved results on the constrained optimization task.


\end{abstract}


\section{Introduction}

{\it Druglikeness} is an essential ingredient for conducting virtual screening of molecules in early-stage drug discovery--prior to synthesis and testing \cite{Lyne2002StructurebasedVS}. 

\begin{figure}[H] 
    \centering
    \includegraphics[width=0.55\textwidth]{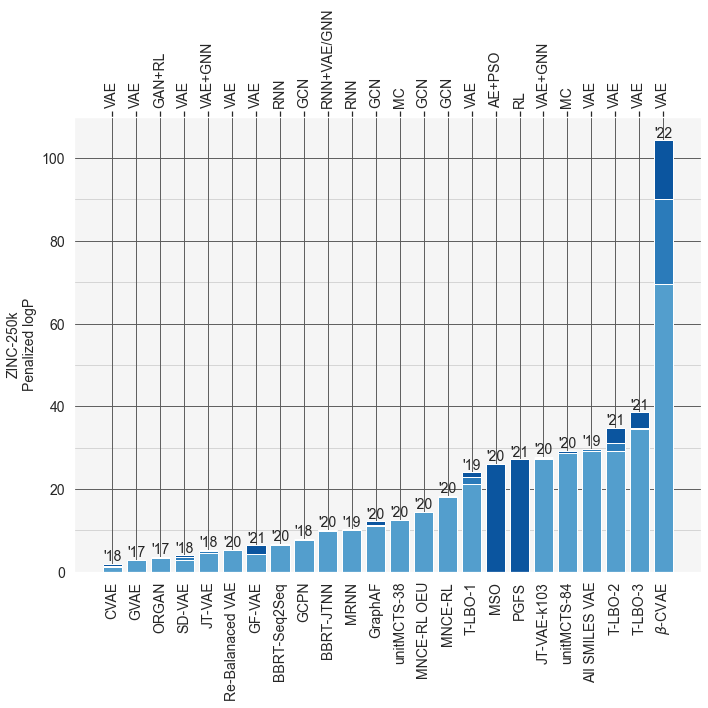}
    \caption{Presented here is a survey of molecular generation studies which utilize the ZINC-250k benchmark dataset. Techniques fall into one of several families, which includes variational autoencoders (VAE), generative adversarial networks (GAN), recurrent neural networks (RNN), graph convolutional networks (GCN), monte carlo (MC), reinforcement learning (RL), and combinations of multiple methods (indicated by a ``+"). We compare these against our own method proposed in this work ($\beta$-CVAE) which make use of conditional variational autoencoders to achieve state-of-the-art performance by a large margin. Top-3 best scores for each method (ranked by pLogP) are presented.}
    \label{fig:main_results}
\end{figure}

Factors such as solubility in fat and water determine if an orally administered drug can both reach and penetrate the cell membrane and are often described by molecular properties such as the octanol-water partition coefficient (Log P) and the quantitative estimate of drug-likeness (QED). While such properties can be measured experimentally they can also be predicted computationally, allowing for brute-force search of candidate molecules. Increasingly, the latest models and techniques from the field of artificial intelligence are being brought to bear on the challenges of drug development, enabling quicker, cheaper, and more effective means of discovery \cite{MaiaEtAl2020}.

The fundamental objective of \textit{de novo} drug synthesis is to uncover molecular structures with desired properties of interest. It has been estimated that the number of realistic drug-like molecules which could ever be synthesized ranges between $10^{23}$ and $10^{60}$ \cite{Polishchuk2013EstimationOT}, immediately highlighting the benefits of favoring intelligent search strategies over brute force ones. The most successful of these intelligent search strategies promises to accelerate the development of novel drugs by significantly decreasing the number of wet-lab experiments needed to discover candidate molecules.

Three categories deep learning model architectures have been consistently explored for \textit{de novo} molecular generation: autoencoders, graph networks, and reinforcement learning. This paper focuses on advancing \textit{autoencoder} based methods, specifically Variational Autoencoders (VAEs). Semantically similar to standard autoencoders, VAEs have been widely used \cite{Bombarelli2018, kusner2017grammar,simonovsky2018graphvae,dai2018syntaxdirected,jin2019junction,Ma2021,yan2020rebalancing, alperstein2019smiles} to \textit{encode} a molecule (usually represented via textual strings) into a lower dimensional latent space, from which it is \textit{decoded} back to the starting molecule. VAEs that learn from molecular strings rely heavily on recurrent modules within the network, such as long short-term memory (LSTM) cells \cite{cheng2016long}, gated recurrent units (GRUs) \cite{cho2014properties}, and self-attention mechanisms \cite{vaswani2017attention}, to learn relationships between user-defined ``tokens", which can be atoms, bonds, rings, or any combination of substructures or functional groups. Recurrent modules, model architecture design principles, and training protocols used in these studies are mainly driven by advancements in the field of Natural Language Processing (NLP) \cite{Payne2020}.

Optimization routines performed after training (\textit{post hoc}) are used to intelligently navigate the learned latent space to optimize some molecular property. Popular choices for these \textit{post hoc} routines include Bayesian Optimization (BO) \cite{kusner2017grammar, jin2019junction} and Particle Swarm Optimization (PSO) \cite{winter2019}. These techniques have been used in both constrained and unconstrained testing environments to maximize properties such as pLogP and QED. Other benchmarks \cite{Brown2019, jin2019junction} gauge how well these optimization routines can re-locate an existing molecule, find molecules with diverging similarity metrics, and locate molecules with varying levels of fingerprint similarity.

In this paper, we advance and simplify existing techniques to condition the latent space of a VAE used for molecular generation and prove that a well-conditioned latent space facilitates \textit{post hoc} molecular optimization for maximizing  pLogP and QED scores. Furthermore, we develop a robust training protocol which automatically adjusts the KL divergence loss weight to maximize the posterior mutual information. We analyze common constraints and limitations from previously published textual based (e.g. SMILES) VAEs for maximizing the pLogP score and present simple solutions for overcoming these deficiencies. As a product of these changes and enhancements, we produce SOTA results for the ZINC-250k unconstrained pLogP benchmark, while closely matching SOTA for the random generation (Gaussian prior) benchmark's validity, novelty, and uniqueness scores, and produce a new SOTA for validity without check portion of the benchmark. Our proposed framework is designed to be customizable so that other NLP architectures can easily supplant the existing encoder and decoder networks. Further, our latent conditioning method is extensible to arbitrary new molecular properties, so that other \textit{de novo} molecular optimization tasks can be studied.

\section{Methodology}

\subsection{Vanilla Variational Autoencoder (VAE) for Sequence Generation} \label{vanilla-vae}

A Variational Autoencoder (VAE) is a generative network derived from a standard autoencoder (AE) framework that observes a space $x$ in terms of a prior distribution over a latent space $p(z)$ and conditional likelihood of generating a data sample from a latent space $p_\phi(x|z)$ \cite{rezende2014stochastic, kingma2014autoencoding, alperstein2019smiles}. The optimal parameter set $\phi^*$ that maximizes the probability of constructing $x$ is given by $\phi^* = \argmax_{x} \sum_{i=1}^{n} \log p_\phi(x_{i})$ \cite{weng2018VAE}. Then the true log likelihood $\log p_\phi(x_{i})$ can be formulated as $\log p_\phi(x_{i}) = \log \int p_\phi(x_{i}|z) p_\phi(z)$. Since this is \textit{intractable}, the evidence lower bound (ELBO) is instead maximized, defined in Eq. \ref{ELBO}. ELBO relies on a variational approximation to the posterior distribution $q_\theta(z|x)$ to maximize the log likelihood.

\begin{equation} \label{ELBO}
\begin{split}
\mathcal{L}(x; \theta, \phi) & =  \mathbb{E}_{q_{\phi}(z|x)}[\log p_{\theta}(x|z)] - D_{\text{KL}}(q_{\phi}(z|x) || p(z)) \\
& \leq \log p(x).
\end{split}
\end{equation}

Following the ELBO formulation, the VAE is comprised of a probabilistic encoder $q_\phi(z|x)$ and decoder  $p_\theta(x|z)$. The encoder is parameterized by $\phi$ and learns a functional mapping between input $x$ to the posterior multivariate Gaussian distribution in the latent space $z$. The decoder, parameterized by $\theta$ learns to reconstruct $x$ from the variational distribution $z$. The VAE seeks to minimize the reconstruction term in ELBO such that the encoder generates meaningful latent vectors which the decoder can subsequently reconstruct. The Kullback–Leibler (KL) loss minimizes the KL divergence between the approximate posterior $q_\theta(z|x)$ and the Gaussian prior distribution $p(z) \sim N(0,I)$. An appealing advantage of the VAE over a standard AE is that it can be used as a \textit{post hoc} generator, capable of decoding lower dimensional latent vectors via a trained decoder.

The vanilla VAE usually suffers from \textit{posterior collapse}, a phenomenon by which the network learns a trivial local optimum of the ELBO objective, causing the variational posterior to be severely misrepresented as the true posterior \cite{yan2020rebalancing, he2019lagging}. \textit{He et al.} \cite{he2019lagging} demonstrate that posterior collapse can be mathematically represented as a local optimum of VAEs whereby both the encoder and decoder networks equivalently represent the latent output $z$ as the prior distribution, shown in Eq. \ref{posterior_collapse}. The authors from \cite{he2019lagging, yan2020rebalancing} show that posterior collapse simply occurs as a direct result of an underrepresented (or \textit{lagging}) reconstruction loss. During training, vanilla (imbalanced) VAEs will equally weight the reconstruction and KL loss terms which leads to the network quickly optimizing the KL term, locating an inaccurate local optimum that poorly approximates the true posterior.

\begin{equation} \label{posterior_collapse}
q_{\phi}(z|x) = p_{\theta}(z|x) = p(z) \mspace{18mu} \forall x \in \R
\end{equation}

Higgins et al. \cite{Higgins2017betaVAELB} introduced the $\beta$-VAE to combat posterior collapse and therefore improve, or disentangle, the latent representation of samples (image-based datasets). The authors \cite{Higgins2017betaVAELB} introduced a hyperparameter $\beta$ in the loss function that is applied \textit{statically} during training, which led to significant improvements in disentanglement scores over the vanilla VAE implementation. The resultant ELBO formulation is shown in Eq. \ref{ELBO_beta}, where $\beta$ is simply a multiplicative weight to the KL loss term. Other works \cite{bowman2016generating} have shown that \textit{dynamically} altering $\beta$ during training can lead to more optimal results. Bowman et al. \cite{bowman2016generating} propose a training scheduler which anneals (linearly increasing, although other approaches such as cyclic ramp and de-ramp have been used) $\beta$ during training and clips it to an empirically derived optimal maximum. Additionally, Yan et al. \cite{yan2020rebalancing} show that this linear annealing process can be effectively used for disentangling latent representations for a $\beta$-VAE that was specifically engineered to generate molecular SMILES strings. Yan et al. \cite{yan2020rebalancing} empirically derive an optimal maximum of $\beta=0.1$ for the ZINC-250k dataset. 

\begin{equation} \label{ELBO_beta}
\begin{split}
\mathcal{L}(x; \theta, \phi) & =  \mathbb{E}_{q_{\phi}(z|x)}[\log p_{\theta}(x|z)] - \beta D_{\text{KL}}(q_{\phi}(z|x) || p(z)) \\
& \leq \log p(x), 0 \leq \beta < 1
\end{split}
\end{equation}

Though the $\beta$-VAE is a robust solution to mitigate posterior collapse and to improve training, molecular generation from such a network can still be limited. Typically, molecular generation tasks are goal-directed, meaning that they involve using a \textit{post hoc} optimization framework which targets specific molecular properties known \textit{a priori}. Such a framework usually employs an optimizer which intelligently navigates a learned posterior, basing its trajectory on intermediate properties from decoded molecules. Vanilla $\beta$-VAEs lack sufficient network constraints and additional functionality to condition the posterior distribution on such properties. Conditional VAEs (CVAEs) can solve this by generating molecules with specific target properties that are imposed by explicitly defined condition vectors. Prior research, e.g. \textit{Lim et al}'s CVAE \cite{articlecvae}, however shows that CVAEs struggle to disentangle the latent molecular representations (evident by PCA/t-SNE reduction techniques of the posterior), and achieve low molecular validity when decoding. Other approaches like \textit{Gomez-Bombarelli et al.} \cite{Bombarelli2018} reframe the standard VAE architecture to promote more latent space disentangling by additionally regressing targets from latent representations. \textit{Gomez-Bombarelli et al}'s VAE illustrates sufficient latent space separation based on the respective targets, but achieves low reconstruction and validity during decoding, while also struggling to produce molecules with high-valued penalized logP scores, where a vanilla $\beta$-VAE \cite{yan2020rebalancing} demonstrated better results (without imposing any conditional constraints). Other unique conditioning techniques for VAEs (including graph-based approaches) have been proposed \cite{rigoni2020conditional, Kang2018} which demonstrate competitive reconstruction and validity rates, as well as low inference times, but have not been applied to common constrained or unconstrained benchmarks. One of the core methods presented in our paper introduces an alternative to conditional sampling which instead directly encodes property information into the latent representations.

\subsection{Posterior Conditioning Techniques} \label{prob-reg}

We first discuss how probabilistic regression techniques can be used to as a conditioning method, then provide our simplifications and other contributions. Zhao et al. \cite{zhao2019variational} made significant contributions to disentangling the latent space of a standard VAE by conditioning the latent vectors on continuous, task-dependent features.

Rather than sampling from a single Gaussian prior $N(0,I)$, Zhao et al. condition $z$ on a target property $c$, effectively formulating a conditional distribution $p(z|c)$ which captures a property-specific prior on latent representations. The regression network predicting $p(z|c)$ is referred to as a \textit{latent generator} \cite{zhao2019variational}. We are able to disentangle the target property $c$ from the prior through the construction of $p(z|c)$, shown in Eq. \ref{conditional_distro}. Here, altering $c$ induces a change in the mean of the resultant Gaussian distribution. Therefore, continuous targets can be encoded in the mean of the distribution.

\begin{equation} \label{conditional_distro}
p(\boldsymbol{z}|c) \sim \mathcal{N}(\boldsymbol{z};\boldsymbol{u}^{\textbf{T}}c,\sigma^2\textbf{I}), \boldsymbol{u}^{\textbf{T}}\boldsymbol{u}=1
\end{equation}

However, $c$ is paired with and dependent on respective input samples $x$, giving the true posterior the final form of: $p(z, c|x)$. If $c$ is unknown during training or for semi-supervised applications, a \textit{probabilistic regressor} $q_{\zeta}(c|x)$ is required. Zhao et al. realize $q_{\zeta}(c|x)$ with a standard feed-forward network to regress continuous targets $c$ from inputs $x$. $q_{\zeta}(c|x)$ is enforced during training with an additional loss term $\log q_{\zeta}(c|x)$ to minimize the difference between the predicted target and its respective ground truth. Finally, to regulate the conditional distribution $p(z, c|x)$, the KL loss term is altered to instead minimize the KL loss between the conditional prior $p(z|c)$ and the posterior generated by the encoder $p_\phi(z|x)$, shown in Eq. \ref{ELBO_reg}. The conditional component of our $\beta$-CVAE is derived from $p(z, c|x)$ since the approximated posterior is \textit{conditioned} on continuous targets.

\begin{equation} \label{ELBO_reg}
\begin{split}
\mathcal{L}(x, c; \theta, \phi, \zeta) &= \mathbb{E}_{q_{\phi}(z|x)}[\log p_{\theta}(x|z)] \\
& -D_{\text{KL}}(q_{\phi}(z|x) || p(z|c)) \\ 
& -\log q_{\zeta}(c|x) \leq \log p(x).
\end{split}
\end{equation}

For our study where target, $c_{i}$, is known for each molecular input $x_{i}$, we instead opt to simply remove the probabilistic regressor $q_{\zeta}(c|x)$ and latent generator $p(z|c)$, and instead formulate a new multivariate Gaussian prior $p(\hat{c}_{i}) \sim N(\hat{c}_{i},I)$, where $\hat{c}_{i}$ is the standardized target  $\hat{c}_{i} =$ \(\frac{c_{i} - \mu_{c}}{\sigma_{c}}\), while $\mu_{c}$ and $\sigma_{c}$ are the mean and standard deviation computed across all targets respectively. Hence, we aim to minimize the KL divergence between the $q_{\phi}(z|x)$ and $N(\hat{c},I)$, rather than using a zero-mean Gaussian prior ($p(z) \sim N(0,I)$), which effectively minimizes the difference between the mean of the posterior and the respective standardized targets. We refer to this latent disentanglement method as ``explicit latent conditioning” (ELC). ELC enables disentanglement by altering latent vectors without imposing additional networks, losses, or altering GRU (or other recurrent units) states which complicate training and increase the complexity and size of the network \cite{zhao2019variational, Kang2018}. ELC allows us to architect the latent space such that molecules with similar values of $\hat{c}$ fall within close proximity. Further, this disentanglement method can be generalized to \textit{any desired molecular properties} external to pLogP and QED such as optoelectronic properties \cite{RICHARDS202143, StJohn2019, harvard-cep}, chemical reactivity, melting point, and solubility \cite{learning-from-cep, SAJEDIAN2020670}. Although this paper focuses on univariate molecular properties, practitioners often seek to generate molecules with multiple constraints or desired ranges of multiple properties. ELC can be easily extended to \textit{multivariate} cases to facilitate multi-property constrained optimization tasks.

We show in Sec. \ref{unconst-opt} that disentangling the latent space with this technique dramatically enhances unconstrained optimization performance. Our ELBO objective function has the final form shown in Eq. \ref{eq:ELBO_final}

\begin{equation} \label{eq:ELBO_final}
\begin{split}
\mathcal{L}(x; \theta, \phi) & =  \mathbb{E}_{q_{\phi}(z|x)}[\log p_{\theta}(x|z)] - \beta D_{\text{KL}}(q_{\phi}(z|x) || p(\hat{c})) \\
& \leq \log p(x)
\end{split}
\end{equation}

\subsection{Mutual Information} \label{mutu_info}

While several dynamic $\beta$ schedulers have been introduced in previous works to regulate the fractional impact of the KL loss term to avoid posterior collapse, we still require a metric for monitoring whether posterior collapse is occurring. Hoffman and Johnson \cite{hoffmanelbonodate} derive the Mutual Information (MI) of a VAE, shown in Eq. \ref{KL_mutual_info_monte_carlo}

\begin{equation} \label{KL_mutual_info_monte_carlo}
\begin{split}
 I_q = & \mathbb{E}_{p_d{(x)}}[D_{\text{KL}}(q_{\phi}(z|x) || p(z))] - \\
& D_{\text{KL}}(q_{\phi}(z) || p(z)) ,
\end{split}
\end{equation}

Yan et al. showed that $I_q\sim0$ is indicative of posterior collapse in a VAE, while their well-trained network achieves a maximum of $I_q\sim4.8$. We use MI as a metric for both monitoring the quality of latent space representations and as a mechanism for deriving the best value for $\beta$ at each training step. We use the same method as \cite{yan2020rebalancing, hoffmanelbonodate} for computing mutual information.

\subsection{MI-centric Training Protocol} \label{mi-training}

After a thorough assessment of prior work \cite{yan2020rebalancing}, it is evident that the pivotal mechanism behind successfully training any VAE for molecular generation is the ability to dynamically alter the KL-loss weight, $\beta$. Further, properly balancing the KL loss during training such that the posterior mutual information is maximized results in higher chemical validity, novelty, and uniqueness during the decoding phase (metrics used in \cite{Zang2020} and presented in Table \ref{tab:zinc-prior}). 

We find that widely used scheduling methods \cite{yan2020rebalancing, bowman2016generating} for adjusting $\beta$ during training are suboptimal as their rigid design prevents the posterior MI from being maximized. Further, the selection of parameters used in the chosen scheduling method (ramp, de-ramp, cyclic timing, among other hyperparamters) is crucial to optimizing network performance. Optimal hyperparamter values are difficult to approximate, especially when altering major architectural features and loss functions, while executing a robust search algorithm to optimize can be compute and time intensive. We instead opt for a more versatile and dynamic approach which involves using a non-linear proportional-integral (PI) controller to alter $\beta$ during training such that the posterior MI reaches a targeted value known \textit{a priori}. \textit{Shao et al.} \cite{shao2020controlvae} present such a method, which approximates the ideal value for $\beta$ at each training step following Eq. \ref{eq:vae-lang}.

\begin{equation}\label{eq:vae-lang}
\vspace{-0.02in}
\beta(t) = \frac{K_p}{1+\exp(e(t))} - K_i \sum_{j=0}^t e(j) + \beta_{min},
\end{equation}

From Eq. \ref{eq:vae-lang}, $e(t)$ is the error function that guides the PI controller at training step $t$ and is originally defined by $e(t) = {v}_{kl} - \hat{v}_{kl}(t)$ \cite{shao2020controlvae}, where $\hat{v}_{kl}(t)$ is the KL loss at the current training step and ${v}_{kl}$ is the target KL loss. Following the empirically derived values from \cite{shao2020controlvae}, we set $K_{p}$ and $K_{i}$ to 0.01 and 0.001, respectively. Targeting an arbitrarily low KL loss leads to naive latent representations and can induce posterior collapse \cite{shao2020controlvae, liu2017unsupervised}, hence careful selection of ${v}_{kl}$ is required. More importantly, an acceptable value for ${v}_{kl}$ does not guarantee maximization of the posterior mutual information, which is our goal. We redefine $e(t)$ to instead target an optimal posterior mutual information score $v_{MI}$. Therefore the PID controller modifies $\beta$ based on the difference between the MI value at each step with the targeted value. However, similar to ${v}_{kl}$ in \cite{shao2020controlvae}, we find that the selection of ${v}_{MI}$ is sensitive, as choosing an arbitrary high value induces a sustained low $\beta$ value that results in poor KL losses. We empirically deduce, aligned with past work \cite{yan2020rebalancing}, that a ${v}_{MI}$ of $\sim4.85$ leads to optimal and stable training results. Following previously described schedulers, we initialize $\beta$ at $0$ to introduce a larger focus on minimizing the reconstruction loss during the beginning epochs of training.

\subsection{$\beta$-CVAE Architecture} \label{main_arch}

\begin{figure*}[ht!] 
    \centering
    \frame{\includegraphics[width=10cm, height=10cm]{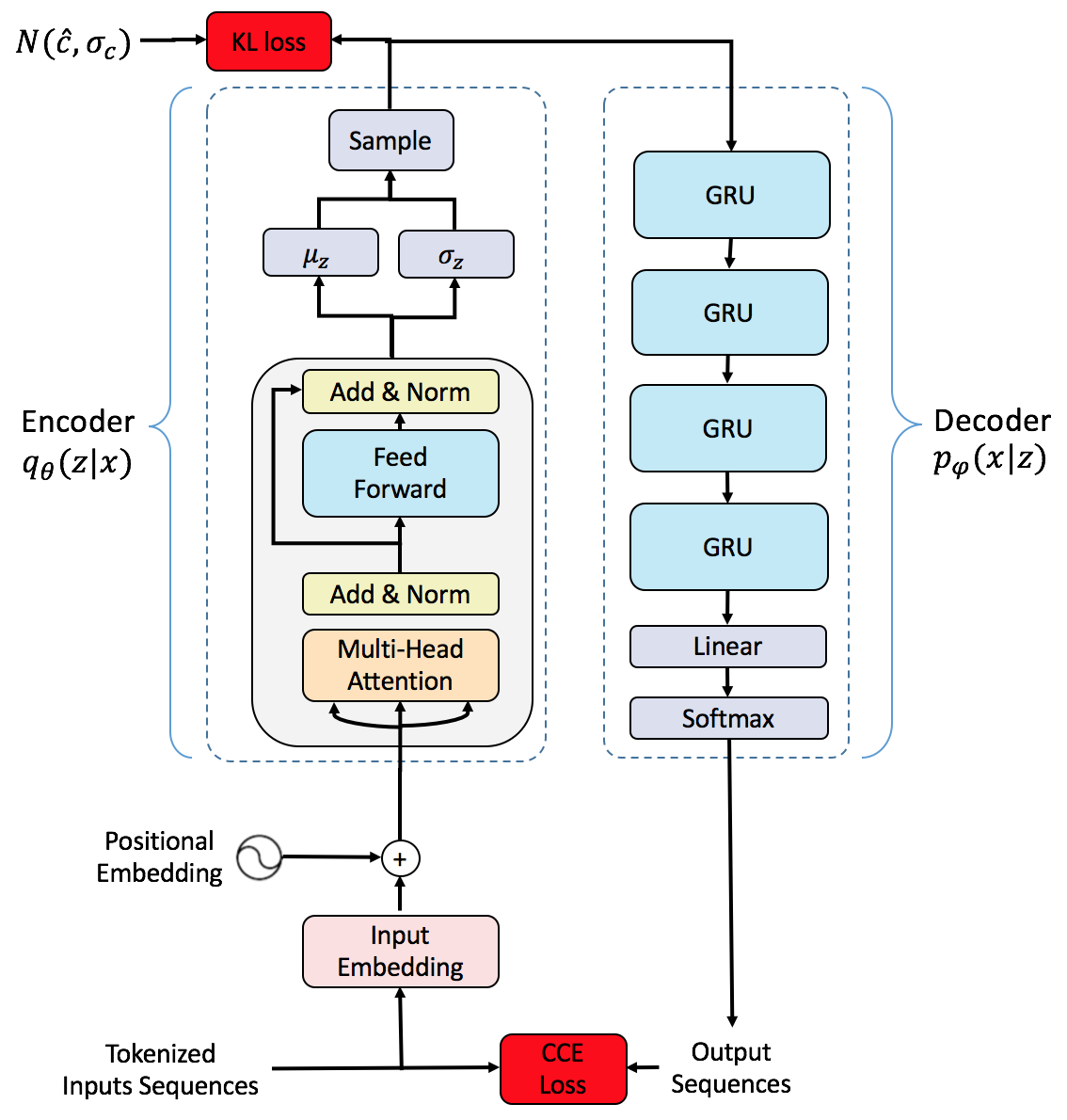}}
    \caption{Main $\beta$-CVAE architecture, which uses a multi-head attention driven encoder network to produce $q_\theta(z|x)$ followed by a GRU-based decoder network that produces $p_\phi(x|z)$. Latent samples produced by the encoder are to compute the KL loss term, while decoder outputs have a softmax activation applied and are passed into a categorical cross-entropy (CCE) loss to drive the decoder outputs to the original tokenized input sequences.}
    \label{fig:arch_main}
\end{figure*}

The final architecture of our proposed $\beta$-CVAE network is illustrated in Fig. \ref{fig:arch_main}, which consists of two major components: an encoder and decoder. For the \textit{encoder}, we transform each tokenized input sequence to continuous multi-dimensional vectors via a standard embedding layer, which is further enhanced by using a \textit{positional} embedding layer, encoding the token position in each embedding vector. The remainder of the encoder uses a basic multi-head attention driven framework employed in the vanilla Transformer architecture \cite{vaswani2017attention}. For the \textit{decoder}, we opt to use a similar architecture to the one used in \cite{yan2020rebalancing} which incorporates Gated Recurrent Units (GRUs) \cite{cho2014properties} over multi-head attention layers (used in the Transformer framework). We also improve our training performance by using \textit{teacher forcing} in our decoder network which was proven to lead to more stable training in \cite{yan2020rebalancing}.

\subsection{VAE Sequence Generation Limitations and Remedies}

Previous methods for \textit{de novo} molecular generation which rely on tokenized strings (e.g. SMILES, InChI) have poorly optimized the penalized logP property for both constrained and unconstrained optimization, whereas graph-based methods have generally prevailed. We conjecture that this poor performance can be attributed to two major deficiencies: (1) insufficient memory for generating long sequences and (2) an absence of specific tokens used to represent larger molecules in string format. We elaborate on each deficiency and present solutions to each in the following sections, while we show results of each remedy in later sections.

\subsubsection{Sequence Length Constraints}

The penalized logP metric \cite{jin2019junction} measures the estimated octanol-water partition coefficient (logP) while simultaneously accounting for ring size and synthetic accessibility, shown in Eq. \ref{plogp_target}.

\begin{equation} \label{plogp_target}
 y(m) = logP(m) - SA(m) - cycle(m),
\end{equation}

A more accurate algorithm for estimating logP was introduced by \textit{Wildman and Crippen} \cite{Crippen}. This involves a simple summation of individual logP scores from molecular constituents known \textit{a priori} \cite{Crippen}, given by Eq. \ref{logp_crippen}, where for some molecule $m$, $n_{i}$ is the number of atoms of type $i$ and $a_{i}$ is the logP of type $i$.

\begin{equation} \label{logp_crippen}
 logP(m) = \sum_{i}{n_{i}a_{i}}
\end{equation}

It is evident from Eq. \ref{logp_crippen} that as we continue to append \textit{relevant} constituents (as defined in \cite{Crippen}) the logP score will continue to \textit{increase} accordingly. Hence, to produce molecules with large logP scores a generative model must be able to accomplish two things: (1) identify molecular constituents with positively contributing logP scores, and more importantly, (2) it must have the capacity to produce \textit{larger} molecules (i.e. \textit{longer} sequences). Current best molecular generation networks trained on ZINC-250k for unconstrained penalized logP maximization have demonstrated the capacity to produce large structures \cite{alperstein2019smiles, grosnit2021highdimensional, rajasekar2020goal, tripp2020sampleefficient}, whereas traditional VAE based networks have been limited to much smaller structures \cite{yan2020rebalancing, jin2019junction, Bombarelli2018}.

NLP-based sequence generation models--relying on functional components like multi-head attention, LSTMs, or GRUs--which do \textit{not} make training protocol changes or incorporate syntax-awareness are typically incapable of producing output lengths much larger than their input lengths. As a result, vanilla VAEs used for SMILES sequence generation are inherently limited by the \textit{absence} of longer sequences in the training dataset.

\begin{figure*}[ht!]
\centering
\subfigure[Training and validation CCE (reconstruction, or log likelihood) loss per epoch.]{
\includegraphics[width=8cm, height=5cm]{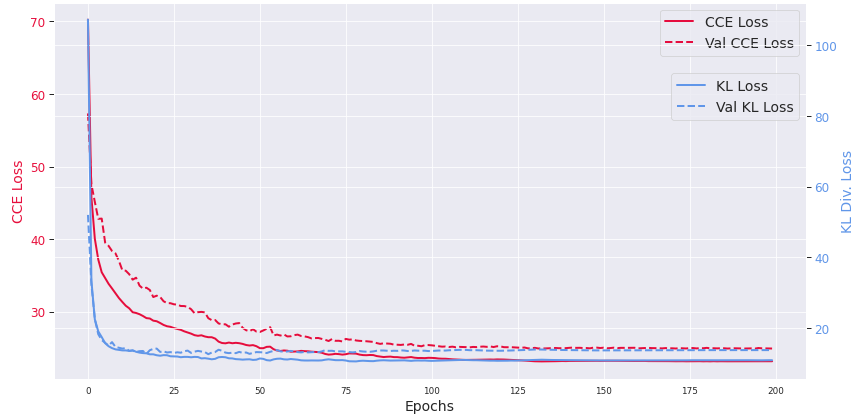}}
\label{fig:cce_loss}
\qquad
\subfigure[Training and validation posterior Mutual Information per epoch.]{
\includegraphics[width=8cm, height=5cm]{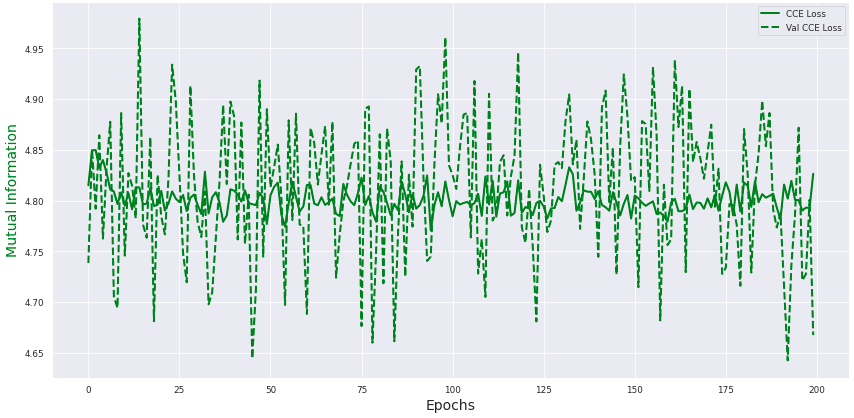}}
\label{fig:mi}
\caption{Training and validation losses and metrics collected over 200 epochs.}
\end{figure*}

\subsubsection{Insufficient Token Representation} \label{known-constraints}

In addition to inadequacies in sequence length the second major limiting factor in achieving higher pLogP scores with VAEs is the insufficiency of tokens present in the initial text corpus. In order to generate molecules with different geometries, number of rings, and bond types, the associated SMILES tokens must be present in the embedding vocabulary so that the network can appropriately utilize them. The clearest example of how this can handicap a model's ability to generate new molecules is the lack of numerical tokens (e.g. 1, 2, 3) which denote a particular ring. The highest ring number in the ZINC-250k dataset is $8$, meaning that only $8$ unique rings can be formed within the entire molecule. Current best molecular generation networks \cite{alperstein2019smiles, grosnit2021highdimensional, rajasekar2020goal, tripp2020sampleefficient} have consistently shown that molecules with the highest pLogP scores consist of multiple concatenated rings (large than $8$). As a result we hypothesize here that a more diverse molecular corpus should assist us in achieving higher quality predictions using our proposed model.

\subsubsection{Dataset Expansion for Overcoming Deficiencies}

We remedy the deficiencies described above by marginally expanding the ZINC-250k dataset to include samples with longer sequences. We first create a library of substructures by aggregating the lowest pLogP scores within the dataset. Following this, we generate $\sim5000$ new samples by randomly appending these substructures to existing molecules at random (syntactically correct) locations. We ensure that the pLogP scores of the new larger molecules do \textit{not} exceed the highest  pLogP score in the ZINC-250k dataset overall. In other words, we are careful not to allow our generative model to benefit from newly created data with inherently better properties. Not only does this expanded corpus provide the model with longer molecules, but it also provides new tokens which can be used to represent new types of structures. We emphasize here that this expanded dataset is only used to train our model on the unconstrained optimization task, while the constrained optimization task (discussed in Sec \ref{unconst-opt}) uses the original ZINC-250k dataset.

\section{Results}
\subsection{Dataset}
We use the ZINC-250k dataset \cite{kusner2017grammar} for all of our experiments. ZINC-250k was constructed by taking $250,000$ random samples from the ZINC database \cite{Bombarelli2018}.

\subsection{Network Training}
All networks described in this paper were implemented using Keras \cite{chollet2015keras} and TensorFlow \cite{abadi2016tensorflow}. All training was performed on an NVIDIA GeForce RTX 2070 GPU, with 8GB of memory.

\begin{figure*}[ht!]
\centering
\subfigure[\textit{No} partial SMILES parser (validity check) to guide the decoding process.]{
\includegraphics[width=8cm, height=7cm]{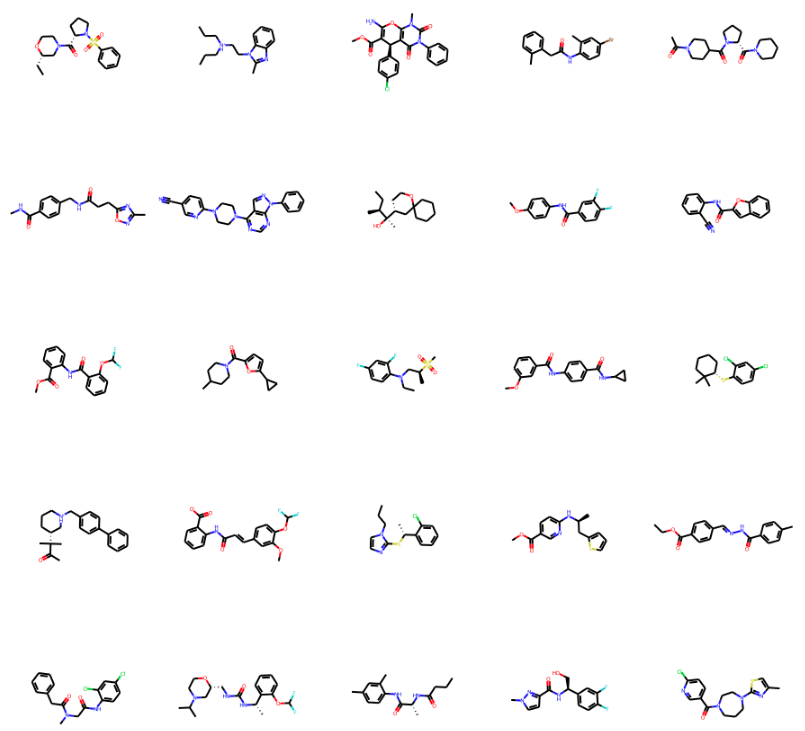}}
\qquad
\subfigure[With partial SMILES parser]{\includegraphics[width=8cm, height=7cm]{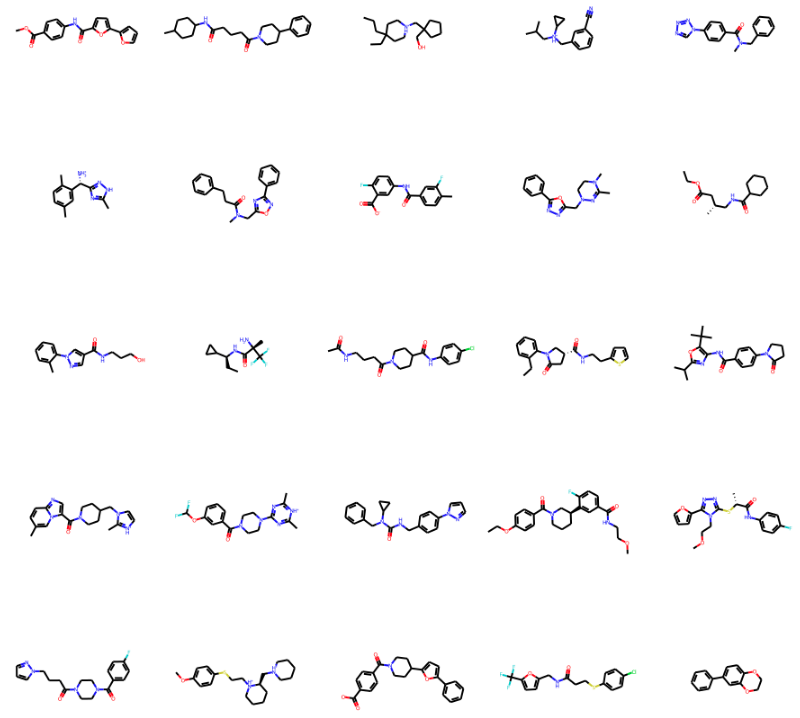}}
\caption{Molecules generated by randomly sampling $N(0,I)$ and decoding.}
\label{fig:losses}
\end{figure*}

We train the $\beta$-CVAE model using separate Adam optimizers \cite{kingma2017adam} for the encoder and decoder networks, starting with a learning rate (LR) of $3\times 10^{-3}$ and $1\times 10^{-3}$, respectively. We employ LR decay callbacks such that when a plateau is detected (no CCE loss or KL loss improvements over $3$ epochs) then the respective LR is decreased by a factor of $0.8$. Each network (one trained for each optimization task) is trained for 200 epochs. The training and validation losses (CCE and KL-divergence) are shown in Fig. \ref{fig:losses} (a). Both losses indicate stable training. We also show the mutual information recorded per epoch in Fig. \ref{fig:losses} (b). The training protocol we introduce in Sec. \ref{mi-training} demonstrates its effectiveness in keeping the mutual information balanced during training.

\subsection{Sampling from Gaussian Prior: Generation and Reconstruction} \label{prior-distro}

\begin{table*}[htb!] 
\small
\centering
\caption{Generation performance on ZINC250K dataset sampling from a normal distribution ($N(0,I)$). * indicates that we instead sampled from $N(\mathcal{U}(-1,1),I)$. For our $\beta$-CVAE networks, we condition on penalized logP scores.}
\vspace{-0.1in}
\begin{tabular}{l c c c c c }
\toprule
 & \textbf{\% Validity} & \textbf{\% Validity w/o check} & \textbf{\% Uniqueness} &  \textbf{\% Novelty}   & \textbf{\% N.U.V.}  \\
\midrule

CVAE \cite{Bombarelli2018} & $0.7$  & N/A  &  N/A & N/A & N/A \\ 
GVAE \cite{kusner2017grammar} & $7.2$  & N/A  &  N/A & N/A & N/A \\
GraphVAE \cite{simonovsky2018graphvae} & $13.5$  & N/A  &  N/A & N/A & N/A \\ 
SD-VAE \cite{dai2018syntaxdirected} & $43.5$  & N/A  &  N/A & N/A & N/A \\ 
JT-VAE \cite{jin2019junction} & $\mathbf{100}$  & N/A  & $\mathbf{100}$ & $\mathbf{100}$ & $\mathbf{100}$ \\
GCPN \cite{you2019graph} & $100$  & $20$  &  $99.97$& $100$    & $99.97$ \\
MRNN \cite{popova2019molecularrnn}& $100$  & $65$  &  $99.89$ & $100$ & $99.89$ \\ 
GraphNVP \cite{madhawa2019graphnvp} & $42.6\pm1.6$  & N/A  &  $94.8\pm0.6$& $100$ & $40.38$ \\ 
GRF \cite{honda2019graph} & $73.4\pm 0.62$  & N/A  &  $53.7\pm 2.13$& $100$    & $39.42$ \\
All SMILES VAE \cite{alperstein2019smiles}& $98.5\pm0.1$ & N/A & $100.00\pm0.00$ & $99.96$ & N/A \\
GraphAF \cite{shi2020graphaf}& $100$ & $68$& $99.10$ & $100$    & $99.10$ \\ 
MoFlow \cite{Zang2020} & $100.00\pm0.00$ &  $\mathbf{81.76\pm0.21}$ &  $\mathbf{99.99\pm0.01}$ & $\mathbf{100.00\pm0.00}$  & $\mathbf{99.99\pm0.01}$ \\

\textbf{GF-VAE} \cite{Ma2021} & $\textbf{100.00}$ &  N/A & $\mathbf{100.00}$ & $\mathbf{100.00}$  & $\mathbf{100.00}$ \\

\midrule
\textbf{$\beta$-VAE (Our)} & $\mathbf{98.77\pm0.13}$ &  $\mathbf{98.28\pm0.06}$ &  $\mathbf{98.31\pm0.15}$ & $\mathbf{99.75\pm0.13}$  & $\mathbf{97.62\pm 0.35}$ \\

\textbf{$\beta$-CVAE (Our)} & $\mathbf{99.64\pm0.08}$ &  $\mathbf{99.44\pm0.05}$ &  $88.69\pm0.26$ & $\mathbf{99.42\pm0.04}$  & $80.11\pm0.36$ \\ 

\textbf{$\beta$-CVAE* (Our)} & $\mathbf{99.63\pm0.04}$ &  $\mathbf{99.48\pm0.07}$ &  $93.83\pm0.26$ & $\mathbf{99.35\pm0.08}$  & $88.40\pm0.45$ \\ 

\bottomrule 
\end{tabular}
\label{tab:zinc-prior}
\end{table*}

\begin{figure*}[t!]
\centering
\subfigure{\includegraphics[width=5cm, height=5cm]{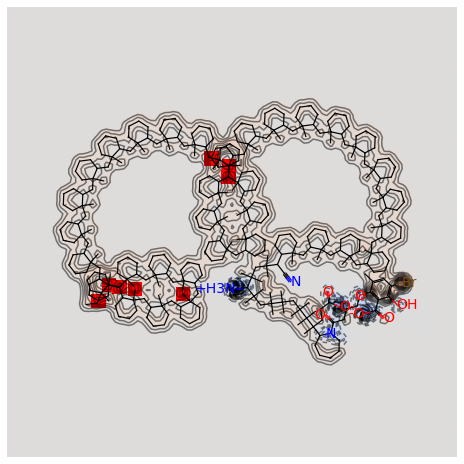}}
\qquad
\subfigure{\includegraphics[width=5cm, height=5cm]{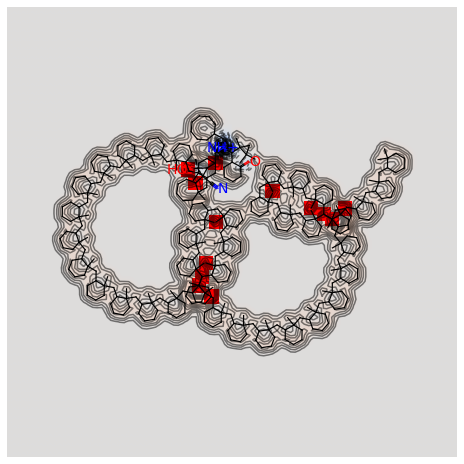}}
\qquad
\subfigure{\includegraphics[width=5cm, height=5cm]{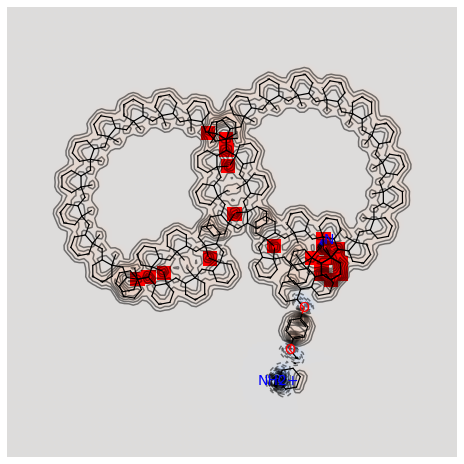}}
\caption{Top-3 molecules generated from our $\beta$-CVAE with the highest pLogP scores. Each molecule has a heatmap overlay which colorizes the atomic contributions to the LogP (non-penalized) score, cycling from red to blue which indicates the highest and lowest contributions respectively.}
\label{fig:top3crippen}
\end{figure*}

Following \cite{jin2019junction}, we evaluate the ability of our $\beta$-CVAE and non-conditional $\beta$-VAE models to generate valid structures strictly from a Gaussian prior $N(0,I)$. Following \cite{Zang2020}, we also report model performance on the following metrics: \textbf{validity} (\% chemically valid), \textbf{uniqueness} (\% unique and valid), \textbf{novelty} (\% falling outside the training dataset), and \textbf{N.U.V.} (\% novel, unique, and valid). 

During the decoding process, we opt to use the partial SMILES parser \cite{partialsmiles}, which checks the syntax of incomplete (partial) SMILES strings, the ability to kekulize aromatic systems, among other checks. We use this parser to guide the decoding process such that if an invalid (intermediate) SMILES string is detected with the network's current prediction of the next token in the decoded sequence, then we \textit{instead} select the next most probable token. This selection process is repeated until either a valid partial SMILES string is found or all tokens have been exhausted. We find that using the partial SMILES parser improves decoding performance. However, to be consistent with \cite{Zang2020}, we compute a final metric that measures the validity without performing any intermediate corrections, denoted as \textbf{validity w/o check}. To compute this, we use a standard decoding process that does not employ the partial SMILES parser or any other intermediate syntax checks.

To remain consistent with \cite{Zang2020, shi2020graphaf}, we sample $10,000$ latent vectors from $N(0,I)$ and decode over $5$ separate runs and record the mean and standard deviation across all runs for each metric in Table \ref{tab:zinc-prior}. Notably, we set a new SOTA in the \textit{validity w/o check} metric at $98.28\pm0.06\%$. Across all other metrics our $\beta$-CVAE and $\beta$-VAE remain competitive with other models,  falling short of the best scores in most cases by less than a few percentage points.

Additionally, we test our $\beta$-CVAE by sampling from $N(\mathcal{U}(-1,1),I)$ to demonstrate that varying the mean of the Gaussian prior (on which pLogP is conditioned) induces more diverse molecules, as only sampling from $N(0,I)$ will generate molecules with an approximated pLogP score of $0$. This is evidenced by the increase in Uniqueness score from $88.69\%$ to $93.83\%$ and N.U.V. score from $80.11\%$ to $88.40\%$.

\subsection{Unconstrained Optimization} \label{unconst-opt}

We follow the JT-VAE paper \cite{jin2019junction} by performing an unconstrained optimization benchmark on the ZINC-250k dataset to maximize the pLogP and Quantitative Estimate of Drug-likeness (QED) \cite{Bickerton2012} scores.

Since the $\beta$-CVAE is conditioned on specific molecular properties, we train two separate networks: one conditioned on pLogP scores, and the other on QED scores. As discussed in Sec. \ref{prob-reg}, the properties must be \textit{standardized} to create a constrained Gaussian prior $p(\hat{c}_{i}) \sim N(\hat{c}_{i},I)$. We note here that for the pLogP network we employ a simple standardization function $\hat{c}_{plogp} =$ \(\frac{c_{plogp} - \mu_{c}}{\sigma_{c}}\). However, since the QED scores have a smaller range of $[0,1]$, we empirically determine that scaling QED scores using a quantile transform to follow a normal distribution, denoted as $Q_{N(0,I)}(x)$, produce better results: $\hat{c}_{QED} = Q_{N(0,I)}(c_{QED})$ .

The unconstrained optimization benchmark seeks to maximize both the plogP and QED scores. We adopt the Molecular Swarm Optimization (MSO) \cite{winter2019} framework for posterior optimization of single properties, as this has demonstrated better results and reduced search latency than the widely used method of coupling a sparse Gaussian process (SGP) with Bayesian Optimization (BO) \cite{jin2019junction}. We initialize the optimizer at $\sim2000$ randomly selected molecules. We execute the PSO routine for $\sim1000$ steps. Table \ref{tab:property_score} shows the results of the benchmark, with our $\beta$-CVAE outperforming the top-3 molecules of the current SOTA method for maximizing pLogP scores, while also matching the highest QED scores. We showcase the molecular structures of the top-3 generated molecules with the highest pLogP scores in Fig. \ref{fig:top3crippen}. Further, we overlay heatmaps onto each of the molecular structures to visualize the atomic contributions to the non-penalized LogP scores.

\begin{table*}[htb!]
\begin{center}
\begin{tabular}{ccccccccc}
\toprule
\multirow{2}{*}{Method} &
\multicolumn{3}{c}{Penalized logP} &
\multicolumn{3}{c}{QED} \\
& 1st & 2nd & 3rd & 1st & 2nd & 3rd \\

\midrule

ZINC (Dataset) & 4.52 & 4.30 & 4.23 & 0.948 & 0.948 & 0.948 \\
\midrule

CVAE \cite{Bombarelli2018} & 1.98 & 1.42 & 1.19 & N/A & N/A & N/A \\

GVAE \cite{kusner2017grammar} & 2.94 & 2.89 & 2.80 & N/A & N/A & N/A \\

ORGAN \cite{guimaraes2018objectivereinforced, damani2019black} & 3.63 & 3.49 & 3.44 & 0.896 & 0.824 & 0.820\\

SD-VAE \cite{dai2018syntaxdirected} & 4.04 & 3.50 & 2.96 & N/A & N/A & N/A\\

JT-VAE \cite{jin2019junction} & 5.30 & 4.93 & 4.49  &
      0.925 & 0.911 & 0.910 \\

Re-Balanaced VAE \cite{yan2020rebalancing} & 5.32 & 5.28 & 5.23 & N/A & N/A & N/A \\

GF-VAE \cite{Ma2021} & 6.47 & 4.42 & 4.42 &
     \textbf{0.948} & \textbf{0.948} & \textbf{0.948} \\

BBRT-Seq2Seq \cite{damani2019black} & 6.74 & 6.47 & 6.42 &
     \textbf{0.948} & \textbf{0.948} & \textbf{0.948} \\

GCPN \cite{you2019graph} & 7.98 & 7.85 & 7.80 &
      \textbf{0.948} & 0.947 & 0.946 \\
     
BBRT-JTNN \cite{damani2019black} & 10.13 & 10.10 & 9.91 &
     \textbf{0.948} & \textbf{0.948} & \textbf{0.948} \\

MRNN \cite{popova2019molecularrnn} & 10.34 & 10.19 & 10.14 & \textbf{0.948} & \textbf{0.948} & 0.947 \\

GraphAF \cite{shi2020graphaf} & 12.23 & 11.29 & 11.05  &
     \textbf{0.948} & \textbf{0.948} & 0.947 \\
     
unitMCTS-38 \cite{rajasekar2020goal} & 12.63 & 12.6 & 12.55 & \textbf{0.948} & \textbf{0.948} & \textbf{0.948} \\

MNCE-RL OEU \cite{xu2020reinforced} & 14.49 & 14.44 & 14.36 & \textbf{0.948} & \textbf{0.948} & \textbf{0.948} \\
     
MNCE-RL \cite{xu2020reinforced} & 18.33 & 18.18 & 18.16  & \textbf{0.948} & \textbf{0.948} & \textbf{0.948} \\

T-LBO-1 \cite{grosnit2021highdimensional} & 24.06 & 22.84 & 21.26 & N/A & N/A & N/A \\

MSO \cite{winter2019} & 26.10 & N/A & N/A  & \textbf{0.948} & N/A & N/A  \\

PGFS \cite{gottipati2020learning} & 27.22 & N/A & N/A & \textbf{0.948} & N/A & N/A \\

JT-VAE ($k=10^{-3}$ retraining \cite{tripp2020sampleefficient}) & 27.84 & 27.59 & 27.21 & N/A & N/A & N/A \\

unitMCTS-84 \cite{rajasekar2020goal} & 29.20 & 28.76 & 28.73 & \textbf{0.948} & \textbf{0.948} & \textbf{0.948} \\

All SMILES VAE (KL-unscaled) \cite{alperstein2019smiles} & 29.80 & 29.76 & 29.11 &
     \textbf{0.948} & \textbf{0.948} & \textbf{0.948} \\

T-LBO-2 \cite{grosnit2021highdimensional} & 34.83 & 31.1 & 29.21 & N/A & N/A & N/A \\

T-LBO-3 \cite{grosnit2021highdimensional} & 38.57 & 34.83 & 34.63 & N/A & N/A & N/A\\
    
\midrule
$\beta$-VAE (Our) & 6.53 & 6.47 & 6.44 & \textbf{0.948} & \textbf{0.948} & \textbf{0.948} \\
 
\textbf{$\beta$-CVAE (Our)} & \textbf{104.29} & \textbf{90.12} & \textbf{69.68} &
     \textbf{0.948} & \textbf{0.948} & \textbf{0.948} \\
\bottomrule
\end{tabular}
\end{center}
\caption{Comparison of the top 3 property scores of generated molecules for both penalized logP and QED scores. "N/A" indicates that no results exist from the original paper.}
\vspace{-5pt}
\label{tab:property_score}
\end{table*}

\subsection{Constrained Optimization}

Similar to Sec. \ref{unconst-opt}, we follow the JT-VAE \cite{jin2019junction} constrained optimization benchmark setup. The 800 molecules with the lowest pLogP scores are selected from the original \textit{test} set as starting molecules to be modified. \textit{Improvement} is measured by finding the increase in pLogP scores between the modification with the highest score and the respective original molecule. \textit{Similarity} is measured by computing the Tanimoto similarity between each modification the respective starting molecule. \textit{Success} is measured by the percent of successful modification, i.e. the percent of modifications that have exceeded the scores of their respective starting molecules.

For this benchmark we choose to disable the conditional sampling portion of our network, and instead train a separate $\beta$-VAE that simply seeks to minimize the KL divergence between $q_\theta(z|x)$ and a zero-mean Gaussian prior $p(z) \sim N(0,I)$, but still utilizes the training protocol outlined in Sec. \ref{mi-training}. The constrained optimization task demands that the network optimize the pLogP scores while remaining within a predetermined similarity boundary $\delta$. Though introducing the conditional sampling mechanism into the network helps us dramatically improve on the unconstrained optimization benchmark, we find that it is \textit{not} as advantageous for the constrained benchmark. Since molecules with high Tanimoto similarity can have large differences in pLogP scores, we hypothesize that conditioning on pLogP rearranges the latent space in such a way that molecules with similar semantic structure are pushed far apart in latent space, which is sub-optimal for the constrained optimization problem.

MSO is run for a maximum of 100 steps on each of the starting molecules. We use 3 swarms each with 450 particles. The optimizer starting point is critical to establish as it can dramatically improve the final improvement of the molecule and reduce search latency. We choose starting points for each molecule $m_{i}$ by encoding the initial molecules to their respective latent vectors $z_{i}$ using our trained encoder $q_\theta(x|z)$, and apply $N(\mu=0,\sigma=1.0)$ to each vector to produce semantically similar neighbors $\hat{z}_{i}$. Since molecules with similar structures (semantic similarities) should be within close proximity to one another in latent space, we can assist the constrained optimization by using this method to locate initial molecules with high Tanimoto similarity to $z_{i}$. We find that this approach produces a more robust starting point for the MSO routine to operate, while choosing a $\sigma$ that is too high can negatively influence results or significantly increase optimization time.

\begin{table}[h]
	\begin{tabular}{lccc}
		\cmidrule(lr){1-4}
		\multirow{3}{*}{Method} &\multicolumn{3}{c}{$\delta=0.4$} \\
		\cmidrule(lr){2-4}
		& Improvement & Similarity & Success \\
		\cmidrule(lr){1-1}\cmidrule(lr){2-4}
		JT-VAE \cite{jin2019junction} & $0.84\pm 1.45$ & {$ 0.51\pm 0.10$} & 83.6\%\\
		MHG-VAE \cite{kajino2019molecular} & $1.00\pm1.87$  & $\bf 0.52\pm\bf 0.11$ & 43.5\%\\
		GCPN \cite{you2019graph} & $2.49\pm1.30$  & $0.47\pm0.08$ & {\bf 100\%}\\
		Mol-CycleGAN \cite{Maziarka2020} & $2.89\pm2.08$ & $\bf 0.52\pm\bf 0.10$ & 58.75\% \\
		MolDQN-naive \cite{Zhou2018} & $3.13\pm1.57$ & N/A & {\bf 100\%} \\
		MolDQ\-boot. \cite{Zhou2018} & $3.37\pm1.62$ & N/A & {\bf 100\%} \\
		MoFlow \cite{Zang2020} & $4.71\pm4.55$  & $0.61\pm0.18$ & 85.75\% \\
		MNCE-RL \cite{xu2020reinforced} & $5.29\pm1.58$  & $0.45\pm0.05$     & {\bf 100\%} \\
		\cmidrule(lr){1-4}
		\textbf{$\beta$-VAE (Our)} & {$\bf5.67\pm2.05$} & $0.42\pm0.05$ & 98.25\% \\
		\cmidrule(lr){1-4}
	\end{tabular}
    \caption{Constrained penalized logP optimizations results with a Tanimoto similarity threshold of $\delta=0.4$. RDKFingerprints were used to compute the Tanimoto similarity. "N/A" indicates that no results exist from the original paper.}
\end{table}

For each step, we evaluate the fitness of the decoded molecules via the following reward function from \cite{Zhou2018}:

\begin{figure*}[htp!]
\centering
\subfigure[$\beta$-VAE]{\includegraphics[width=8cm, height=8cm]{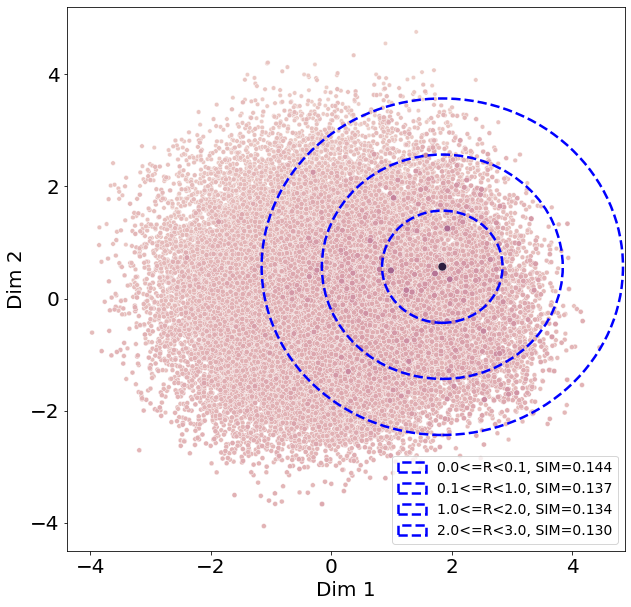}}
\qquad
\subfigure[$\beta$-CVAE (conditioned on pLogP)]{\includegraphics[width=8cm, height=8cm]{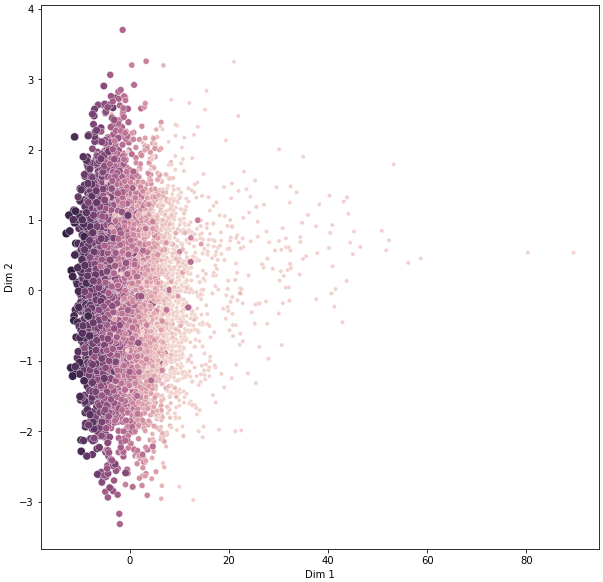}}
\caption{PCA of posterior of the validation set, where each plotted point indicates a projected latent vector of a molecule into 2D space. (a) shows results from the $\beta$-VAE with no latent space conditioning, where the points are colorized based on the Tanimoto similarity between a single candidate molecule (darker indicates higher similarity). (b) shows results from the $\beta$-CVAE conditioned on pLogP, where the points are colorized and sized by the molecule's respective pLogP score (darker and larger indicates a higher pLogP score).}
\label{fig:cvae-pca}
\end{figure*}

\small
\[
\mathcal{R}(s) = \begin{cases}
		\mbox{p}(m_i) - \lambda \times (\delta - \mbox{SIM}(m_{i}, m_0))						 & \mbox{if}\hspace{1em} \mbox{SIM}(m_i, m_0) < \delta \\
		\mbox{p}(m_i)  & \mbox{otherwise}
	\end{cases}
\]

\normalsize
$SIM$ is defined as the Tanimoto similarity between the decoded molecules $m_{i}$ and the candidate molecule $m_{0}$, we choose $\lambda=100$ (same value used in the original work \cite{Zhou2018}), and $\mbox{p}(m)$ indicates computing the pLogP score of $m_i$. We use RDKFringerprints \cite{rdkit}, a ``Daylight-like" substructure fingerprint method, to compute the Tanimoto similarity. We find that this reward function facilitates the search as decoded molecules that have \textit{not} met the Tanimoto threshold are still used to guide the swarm, rather than naively returning a null improvement.

Our $\beta$-VAE model performs very well in the constrained benchmark. In terms of improvement, it outperforms previous methods \cite{jin2019junction, you2019graph, Maziarka2020, kajino2019molecular}, while demonstrating a slightly higher mean improvement than the current SOTA \cite{xu2020reinforced}. Although other methods have demonstrated slightly higher success rates, our method achieves a competitive $98.25\%$. These optimal results further validate our training protocol, as optimally targeting sufficient posterior MI increases the average improvement.

\subsection{Posterior Dimensionality Reduction}

We conduct a two-dimensional (2D) Principal Component Analysis (PCA) on the learned posterior of both the standard $\beta$-VAE and $\beta$-CVAE. For the $\beta$-VAE, this allows us to visualize the Tanimoto similarity between a single candidate molecule and neighboring molecules in latent space. The plots shown in Fig. \ref{fig:cvae-pca} are 2D PCA plots, where each point on the plots represents a projected latent vector from an encoded molecule from the ZINC-250k validation set.

We compute the Tanimoto similarity at several annuli centered around a candidate molecule $m_{i}$. We show that as we traverse the molecular space, moving from more distant annuli ($2<=R<3$) to closer ones ($0.0<=R<0.1$), the average Tanimoto similarity between the $m_{i}$ and all other molecules within each annulus \textit{increases} from $SIM=0.130$ to $SIM=0.144$  respectively. That is, as we approach $m_{i}$ in reduced latent space, neighboring molecules become more structurally similar.

The 2D PCA plot for the conditional VAE is shown in Fig. \ref{fig:cvae-pca} (b). Each point in this space represents a projected latent vector from an encoded molecule, where color is used to represent penalized logP score. As is evident from the plot, we have successfully imposed target-property dependent structure in the latent space (grouping molecules with similar penalized LogP scores); effectively disentangling the latent space. 
\section{Conclusion}

In this paper, we revive a seemingly obsolete technique for \textit{de novo} molecular generation and achieve new SOTA results for several benchmarks. Textual based VAEs have shown little promise when compared to more modern graph and flow based models. However, we show how simple architectural changes, more robust and flexible training protocols, and specific techniques for overcoming dataset limitations can dramatically enhance the performance of VAE models. 

A second, major contribution of this work is to demonstrate how the latent space of VAE models can be purposefully structured to our advantage--either in terms of semantic structure (Tanimoto similarity) or in terms of desired target properties (pLogP). This is achieved by modifying the KL divergence loss to encode directly to the desired molecular property in the posterior. Consequently, we emphasize here that this simple strategy can be extended to guide the molecular optimization for {\it any} desired property (or set of properties) for which we wish to discover new candidate molecules.

\bibliographystyle{unsrt} 
\bibliography{denovo.bib} 

\end{document}